\documentclass[aps,amssymb,pra,showpacs,twocolumn,floatfix]{revtex4}

\usepackage{epsf}
\usepackage{amsmath}
\usepackage{graphicx}
                       
\begin{document}
         
\title {Thread bonds in molecules}

\author{B. Ivlev}

\affiliation
{Instituto de F\'{\i}sica, Universidad Aut\'onoma de San Luis Potos\'{\i},
San Luis Potos\'{\i}, 78000 Mexico\\}

\begin{abstract}

Unusual chemical bonds are proposed. Each bond is almost covalent but is characterized by the thread of a small radius, 
$\sim 0.6\times 10^{-11}{\rm cm}$, between two nuclei in a molecule. The main electron density is concentrated outside the thread as in a 
covalent bond. The thread is formed by the electron wave function which has a tendency to be singular on it. The singularity along the thread 
is cut off by electron ``vibrations'' due to the interaction with zero point electromagnetic oscillations. The electron energy has its typical 
value of $(1-10)~{\rm eV}$. Due to the small thread radius the uncertainty of the electron momentum inside the thread is large resulting in a 
large electron kinetic energy $\sim 1~{\rm MeV}$. This energy is compensated by formation of a potential well due to the reduction of the energy 
of electromagnetic zero point oscillations. This is similar to formation of a negative van der Waals potential. Thread bonds are stable and 
cannot be created or destructed in chemical or optical processes.

\end{abstract} \vskip 1.0cm

\pacs{31.10.+z, 31.15.A-, 03.65.-w}

\maketitle

\section{INTRODUCTION}
\label{intr}
Various types of chemical bonds in molecules are well studied. See for example \cite{PAULING,GRAY,KOCK}. This is a branch of chemical physics 
with the well established set of basic phenomena. Methods of quantum chemistry are effectively applied to develop that field. The common feature
of chemical bonding is the typical energy. Despite a variety of bond types there is the upper limit of energies involved into the phenomena. It 
is the atomic scale energy of the order of ten electron volts. Also the typical spatial scale in chemical bonding is no shorter than the atomic 
size. 

It looks unusual if energies, involved into molecular bonding, are of the order of 1~MeV and the spatial scale is $10^{-3}$ of the size of 
hydrogen atom. At the first sight, it is impossible since chemical bonding is associated with the atomic processes.

The point is that the electron state can have a tendency to be singular along the line (thread) connecting two nuclei in a molecule. The origin
of such singularity is clear from Schr\"{o}dinger equation in cylindrical coordinates which can have the solution logarithmically singular on 
the axis. The electron ``vibrates'' due to the interaction with zero point electromagnetic oscillations and the singularity smears out within 
the thread of a small radius $0.6\times 10^{-11}{\rm cm}$. This radius can be determined accounting for electron-photon interaction in 
quantum electrodynamics since due to short scales the electron momentum becomes relativistic.

The exact state, where the electron is coupled to photons, is steady and is characterized by the certain total energy which is conserved. In 
terms of Schr\"{o}dinger equation it would an energy eigenvalue. The state, considered either far from the thread or inside it, corresponds to 
the same energy. The large contribution to the electron kinetic energy $\sqrt{(mc^2)^2+(\hbar c/r_T)^2}-mc^2\simeq 2.93~{\rm MeV}$ inside the 
thread should be compensated by a reduction of the electromagnetic energy. The latter can be roughly interpreted as a potential well at the 
thread region. One can say that the thread state with the chemical electron energy $(1-10)~{\rm eV}$ corresponds to upper energy states in the 
deep potential well, of the order of 1~MeV, and of the radius $0.6\times 10^{-11}{\rm cm}$. This well is extended along the thread. The electron 
number is determined by the outer region with respect to the thread. This region is of the order of the Bohr radius as in a covalent bond.

This is not the unique situation of application of quantum electrodynamics in chemistry. The famous example is van der Waals forces when quantum 
electrodynamics is involved. In that case the certain attraction potential is also formed due to the reduction of the energy of electromagnetic 
zero point oscillations. 

The high energy parts, involved into thread formation, correspond to the typical time $10^{-22}{\rm s}$. Optical processes are slow compared to 
that time. They lead to an adiabatic motion of the thread parameters but absorption probability of such quanta is exponentially small. The life 
time of the thread bond itself is exponentially large (practically infinity) until a high energy particle or $\gamma$-quantum destroys it. Also 
the thread bond cannot be created in chemical or optical processes. The above high energy impact is necessary for that. 

Thread bonds can be created by an external high energy radiation. One can put a question on a role of such stable bonding in biological processes.

In section \ref{two} the singular (between two Coulomb centers) solution of Schr\"{o}dinger equation is obtained. In section \ref{thread} the
inner thread properties are studied.
\section{Model with two Coulomb centers}
\label{two}
Electrons wave function in a molecule is complicated. The electron motion is collective and is not reduced to a single electron problem. The 
electron wave function is smooth in the space. We focus on the certain specificity of this function. If to consider usual atomic sizes the 
electron wave function can be formally singular on some line connecting two nuclei in the molecule. In the exact wave function the singularity 
is cut off at distances shorter than atomic ones.

Therefore the problem is separated in two parts. First one should find the singular solution of quantum mechanical equations for electrons
in the molecule. Second one should analyze mechanisms of smearing of that singularity. In this section we focus on the singularity formation on 
the line connecting two nuclei which can be treated in the molecule as static ones according to the Born-Oppenheimer approximation. 

Instead of solving the problem in full one can simplify it considering a single electron. The main features of the singular wave function can be 
demonstrated studying the artificial situation when one electron is in the field of two positive charges fixed at two points. The singularity line, 
connecting two nuclei in the molecule, is associated with large energies and is hardly influenced by chemical forces. For this reason the 
singularity line between two real nuclei in the molecule is of the same type as between two point charges formally fixed in the space.
\subsection{Formalism}
\begin{figure}
\includegraphics[width=5cm]{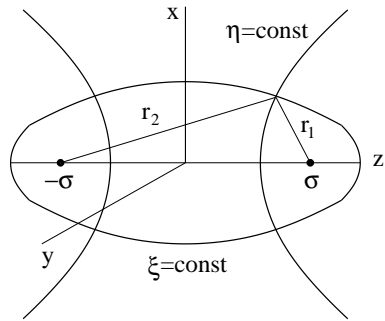}
\caption{\label{fig1} Intersections of surfaces of constant elliptic coordinates $\xi$ and $\eta$ with the plane $y=0$ are shown. Two Coulomb 
centers are at the points $z=\pm\sigma$.}
\end{figure}

To study the electron in the Coulomb field of two positive point charges at the points $z=\pm\sigma$ it is convenient, instead of cylindrical 
coordinates $r=\sqrt{x^2+y^2}$, $z$, and $\varphi$, to use the elliptic ones $\xi$, $\eta$, and 
$\varphi$ \cite{LANDAU4}
\begin{equation}
\label{1}
\xi,\eta=\frac{r_2\pm r_1}{2\sigma}.
\end{equation}
Here $r_{1,2}=\sqrt{(z\mp\sigma)^2+r^2}$ are distances to the Coulomb centers shown in Fig.~\ref{fig1}. The surface of a constant $\xi$ is 
the ellipsoid
\begin{equation}
\label{2}
\frac{z^2}{\sigma^2\xi^2}+\frac{r^2}{\sigma^2(\xi^2-1)}=1
\end{equation}
with the focuses at $z=\pm\sigma$. The surface of a constant $\eta$ is the hyperboloid
\begin{equation}
\label{3}
\frac{z^2}{\sigma^2\eta^2}-\frac{r^2}{\sigma^2(1-\eta^2)}=1
\end{equation}
with the focuses at the same points. The coordinate $\xi$ takes values from $1$ to $\infty$ and $\eta$ from $-1$ to $1$. Intersections of the 
surfaces (\ref{2}) and (\ref{3}) with the plane $y=0$ are shown in Fig.~\ref{fig1}. 

The Coulomb interaction potential is
\begin{equation}
\label{4}
U=-\frac{Ze^2}{r_1}-\frac{Ze^2}{r_2}=-\frac{2Ze^2}{\sigma}\frac{\xi}{\xi^2-\eta^2},
\end{equation}
where $Ze$ is the positive charge at each center. The Schr\"{o}dinger equation for the electron
\begin{equation}
\label{5}
-\frac{\hbar^2}{2m}\nabla^2\psi+U\psi=E\psi
\end{equation}
in elliptic coordinates takes the form
\begin{eqnarray}
\nonumber
&&-\frac{\hbar^2}{2m\sigma^2(\xi^2-\eta^2)}\left[\frac{\partial}{\partial\xi}(\xi^2-1)\frac{\partial\psi}{\partial\xi}
+ \frac{\partial}{\partial\eta}(1-\eta^2)\frac{\partial\psi}{\partial\eta}\right]\\
\label{6}
&&-\frac{2Ze^2}{\sigma}\frac{\xi}{\xi^2-\eta^2}\psi=E\psi.
\end{eqnarray}
We consider an axially symmetric wave function. Since the variables are separated it has the form $\psi(\xi,\eta)=\psi_1(\xi)\psi_2(\eta)$. One 
can introduce dimensionless parameters $\varepsilon=-2m\sigma^2E/\hbar^2$ and $p=4Z\sigma/r_B$ where $r_B=\hbar^2/(me^2)$ is the Bohr radius.

After separation of variables the two Schr\"{o}dinger equations are
\begin{equation}
\label{7}
-\frac{\partial}{\partial\xi}(\xi^2-1)\frac{\partial\psi_1}{\partial\xi}=(p\xi+\beta-\varepsilon\xi^2)\psi_1
\end{equation}
\begin{equation}
\label{8}
-\frac{\partial}{\partial\xi}(1-\eta^2)\frac{\partial\psi_2}{\partial\eta}=(\varepsilon\eta^2-\beta)\psi_2\,,
\end{equation}
where $\beta$ is some constant. 

The variable $\eta$ can be written as $\eta=\cos\chi$. Then Eq.~(\ref{8}) has the form
\begin{equation}
\label{9}
\frac{1}{\sin\chi}\frac{\partial}{\partial\chi}\left(\sin\chi\frac{\partial\psi_2}{\partial\chi}\right)=(\beta-\varepsilon\cos^2\chi)\psi_2.
\end{equation}
The solution, non-singular at $\chi=0$, should be also non-singular after continuation to $\chi=\pi$. This is the condition to choose the 
parameter $\beta$ for a given $\varepsilon$. When two Coulomb centers coincide ($\sigma=0$ and therefore $\varepsilon=0$) that condition turns 
to $\beta=-l(l+1)$ as for Legendre polynomials \cite{LANDAU1}. In that case the variable $\chi$ coincides with the azimuthal angle $\theta$.
\subsection{Close Coulomb centers}
Below two close Coulomb centers are considered under the condition $\sigma\ll r_B$. We study the state which is isotropic ($l=0$) in the limit 
$\sigma=0$. At a finite $\sigma$ there is the small correction to the wave function which can be written as $\psi_2=1+\delta\psi_2$. Using the 
relation
\begin{equation}
\label{10}
\sin\chi\frac{\partial\delta\psi_2}{\partial\chi}=\int d\chi(\beta-\varepsilon\cos^2\chi)\sin\chi
\end{equation}
one can obtain
\begin{equation}
\label{11}
\delta\psi_2=\left(\frac{\varepsilon}{3}-\beta\right)\ln\left(2\cos^2\frac{\chi}{2}\right)-\frac{\varepsilon}{6}\sin^2\chi.
\end{equation}
The solution (\ref{11}) is finite at $\chi=0$ ($\eta=1$). In order to get it finite at $\chi=\pi$ ($\eta=-1$) it should be $\beta=\varepsilon/3$.
Note that $\varepsilon\sim p^2\sim\sigma^2/r^{2}_{B}$ are small.

At $\sigma\ll r_B$ there is a small region of the size $\sigma$ around the centers in Fig.~\ref{fig1}. This region corresponds to 
$\xi\sim\eta\sim 1$. The region of the order of the Bohr radius $r_B$ is much larger and relates to large $\xi$. We consider first this region. 
In Eq.~(\ref{7}) one can omit $\beta$ and to write $\xi^2-1\simeq\xi^2$. In the limit $1\ll\xi$ Eq.~(\ref{7}) takes the form
\begin{equation}
\label{12}
-\frac{\partial^2\psi_1}{\partial\xi^2}-\frac{2}{\xi}\frac{\partial\psi_1}{\partial\xi}-\frac{p}{\xi}\psi_1=-\varepsilon\psi_1,
\end{equation}
which coincides with the radial Schr\"{o}dinger equation with $l=0$ in the Coulomb field of the point charge $2Ze$ \cite{LANDAU1}. The solution 
of (\ref{12}), decaying on infinity and finite at small distances, corresponds to the eigenvalue $\varepsilon=p^2/4$. This value relates to the
ground state energy in the Coulomb field of the point charge $2Ze$.

For our purposes one needs a solution which also decays on infinity but is singular at $r=0$ and $z^2<\sigma^2$. To obtain that one should 
write the solution of (\ref{12}) in the form \cite{LANDAU1}
\begin{eqnarray}
\nonumber
\psi_1(\xi)=\xi^{p/(2\sqrt{\varepsilon})-1}\exp(-\xi\sqrt{\varepsilon})\\
\label{13}
\times G\left(1-\frac{p}{2\sqrt{\varepsilon}},-\frac{p}{2\sqrt{\varepsilon}},-2\xi\sqrt{\varepsilon}\right),
\end{eqnarray}
where 
\begin{equation}
\label{14}
G(\alpha,\beta,v)=1+\frac{\alpha\beta}{1!v}+\frac{\alpha(\alpha+1)\beta(\beta+1)}{2!v^2}+...
\end{equation}
To obtain the solution at shorter distances one should use the asymptotics of the function (\ref{14}) at $v\ll 1$. Thus at small 
$[1-p/(2\sqrt{\varepsilon})]$ one can obtain from Eq.~(\ref{13}) \cite{LANDAU1}
\begin{equation}
\label{15}
\psi_1(\xi)=1+\left(1-\frac{p}{2\sqrt{\varepsilon}}\right)\frac{1}{2\xi\sqrt{\varepsilon}}\,,\hspace{0.3cm}1\ll\xi\ll r_{B}/\sigma.
\end{equation}
On the other hand, at not large $\xi$ the left-hand side of Eq.~(\ref{7}) is the principal one and the solution is
\begin{equation}
\label{16}
\psi_1(\xi)=1+\frac{1}{4\sqrt{\varepsilon}}\left(1-\frac{p}{2\sqrt{\varepsilon}}\right)\ln\frac{\xi+1}{\xi-1},
\hspace{0.3cm}1<\xi\ll r_{B}/\sigma.
\end{equation}
Eq.~(\ref{16}) goes over into the form (\ref{15}) when their applicability intervals overlap. 

The wave function along the line, connecting two Coulomb centers in Fig.~\ref{fig1}, now can be written at $|z^2-\sigma^2|,r^2\ll\sigma^2$ in 
the form 
\begin{eqnarray}
\nonumber
&&\psi(r,z)=1-\frac{r_B}{16Z\sigma}\left(1-\frac{E}{E_0}\right)\\
\label{17}
&&\times\ln\frac{8\sigma^2}{z^2-\sigma^2+\sqrt{(z^2-\sigma^2)^2+4\sigma^2r^2}}\,,
\end{eqnarray}
where $E_0=-m(2Ze^2)^2/(2\hbar^2)$. Eq.~(\ref{17}) is also valid in the vicinity ($r^2\ll(\sigma^2-z^2))$ of the entire line ($z^2<\sigma^2$)
between the centers, where
\begin{equation}
\label{18}
\psi(r,z)=1-\frac{r_B}{8Z\sigma}\left(1-\frac{E}{E_0}\right)\ln\frac{2\sqrt{\sigma^2-z^2}}{r}\,.
\end{equation}
At large distances, as follows from (\ref{13}) - (\ref{14}) and the definition (\ref{1}),
\begin{equation}
\label{19}
\psi(r,z)=R^{\sqrt{E/E_0}-1}\exp\left(-R\frac{2Z}{r_B}\sqrt{\frac{E}{E_0}}\right),\hspace{0.3cm}r_B\ll R,
\end{equation}
where $R^2=r^2+z^2$.

\begin{figure}
\includegraphics[width=5.5cm]{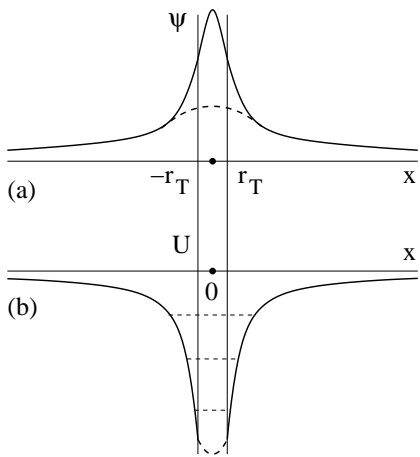}
\caption{\label{fig2} Features of the thread state at $y=0$ and constant $\eta$. Dashed curves correspond to the absence of tread. Thin 
vertical lines restrict the thread region. (a) Thread wave function. (b) Energy levels in the potential well in the absence of thread.}
\end{figure}

The wave function of the electron in the Coulomb field of two positive point charges $Ze$ exponentially decays at large distances (\ref{19}). 
But on the line, connecting two charges, $\psi$  has the logarithmic singularity (\ref{18}) if the energy does not coincide with the eigenvalue 
$E_0$. The absence of singularities is a usual condition to determine an eigenvalue. The eigenvalue $E_0$ coincides with one in the Coulomb 
field of one point charge $2Ze$. In the limit considered, $\sigma\ll r_B$, corrections to that eigenvalue are small. $E_0$ is the lowest energy 
level shown in Fig.~\ref{fig2}(b) where the potential energy (\ref{4}) relates to the curve with the dashed insert. The eigenfunction is plotted 
in Fig.~\ref{fig2}(a) with the dashed part of the curve. 

When the energy differs from $E_0$ the singular wave function is not physical at the first sight. But the situation is more complicated as 
described in Sec.~\ref{thread}.
\section{ELECTRON THREADS}
\label{thread}
The electron wave function, for the artificial case of two static point charges, and the wave function for two partner nuclei in a real molecule 
have common properties. This happens since at short distances (less than $r_B$) atomic forces are not significant. In the real molecule 
close to a nucleus, that is below the shell of inner electrons, the wave function is of the type (\ref{17}). The same is valid for the nucleus 
at a partner site. Between the partner sites the wave function is logarithmically singular, as (\ref{18}), along the connecting line 
$-\sigma<z<\sigma$. In a real molecule $\sigma\sim r_B$. In a close vicinity ($r\ll\sigma$) of the line, connecting two partner sites, the wave 
function, with the logarithmic accuracy, can be written in the form
\begin{equation}
\label{20}
\psi=
\begin{cases}
A(\sigma)\ln[\sqrt{(z-\sigma)^2+r^2}+(z-\sigma)],\hspace{0.4cm}|z-\sigma|\ll\sigma\\
2A(z)\ln r\,,\hspace{3.7cm}\sqrt{\sigma^2-z^2}\sim\sigma \\
A(-\sigma)\ln[\sqrt{(z+\sigma)^2+r^2}-(z+\sigma)],\hspace{0.1cm}|z+\sigma|\ll\sigma
\end{cases}
\end{equation}
Here $A(z)$ is the certain function accounting for real intramolecular forces. The exact form of this function is not crucial for our purposes.

An electron, acted by an external macroscopic field, is spatially smeared due to the interaction with zero point electromagnetic oscillations
\cite{LANDAU3}. If $\vec u$ is the electron displacement from its mean field position then $\langle\vec u\rangle=0$ and 
$r_T=\sqrt{\langle u^2\rangle}$ is finite. In an external field (for example Coulomb one) due to the uncertainty in electron positions it probes 
various parts of the potential. This changes its energy. The famous example of this phenomenon is the Lamb shift of atomic levels \cite{LANDAU3}. 

The electron momentum is not conserved in virtual processes and the electron ``vibrates''. The uncertainty $r_T$ results in small corrections  
(such as Lamb shift) when the spatial scale of the electron wave function is larger than $r_T$. But when the wave function, as a solution of 
quantum mechanical equations, has a singularity that uncertainty plays a significant role resulting in smearing of the singularity within $r_T$.

Therefore the singularity line between two nuclei in a molecule has the finite width of the order of $r_T$. It can be called {\it thread bond}. 
In Fig.~\ref{fig2}(a) the wave function increases towards the tread and it is cut off on the thread size $r_T$. The potential energy in 
Fig.~\ref{fig2}(b) is of the usual Coulomb type outside the thread region. Inside the thread the mean field potential no more exists because 
photon degrees of freedom enter the game. 

The singularity of the wave function is cut off on the distance $r_T$ and the state becomes physical. So the expression (\ref{20}) corresponds 
to not small distances $r^2,|\sigma^2-z^2|$. According to Eqs.~(\ref{17}) and (\ref{18}), the singular solution of the Schr\"{o}dinger 
equation exists at any pre-logarithmic factor $(1-E/E_0)$. This means that the electron energy of the thread state is not quantized due to the 
interaction with photons. This can be explained in another way. Broadening of discrete levels in a well (quasi-continuous density of states) is 
narrow and can be calculated from the perturbation theory. Calculation of smearing of the singular wave function requires all orders of the 
perturbation theory. In this case the density of states is smoothly continuous.
\subsection{Estimate of the thread radius}
Unlike the Lamb shift, it is impossible to apply the perturbation theory to calculate smearing of a singular wave function of electron. In that 
problem one should account for all orders of the perturbation theory. For this reason, we use below the approximate method just to estimate the
electron mean squared displacement \cite{MIGDAL}. The method is successfully applied for study of the Lamb shift. In this method the electron 
motion under the action of zero point oscillations can be described by the equation
\begin{equation}
\label{A1}
m\frac{d^2\vec u}{dt^2}+m\Omega^2\vec u=-e\vec{\cal E},
\end{equation}
where $m$ is the electron mass and $\Omega\sim me^4/\hbar^3$ is the electron rotation frequency in the atom. 

One can use the Fourier expansion 
\begin{equation}
\label{A2}
\vec u(\vec R,t)=\sum_{k}\vec u_{k}\exp(i\vec k\vec R-i\omega_{k}t)
\end{equation}
and analogous one for the fluctuating electric field $\vec{\cal E}(\vec R,t)$. Since $\vec u(\vec R,t)$ is real it should be 
$\vec u^{*}_{k}=\vec u_{-k}$ and $\omega_{-k}=-\omega_{k}$ in the expansion (\ref{A2}). The condition $uk\ll 1$ has to be held in this method.
The solution of Eq.~(\ref{A1}) is of the form
\begin{equation}
\label{A3}
\vec u_{k}=\frac{e\vec{\cal E}_{k}}{2m|\omega_k|}\left(\frac{1}{|\omega_k|+\Omega}+\frac{1}{|\omega_k|-\Omega}\right).
\end{equation}
According to the quantum mechanical approach, Eq.~(\ref{A3}) should be modified as
\begin{equation}
\label{A4}
\vec u_{k}=\frac{e\vec{\cal E}_{k}}{2m|\omega_k|}\left(\frac{\sqrt{1+n_k}}{|\omega_k|+\Omega}+\frac{\sqrt{n_k}}{|\omega_k|-\Omega}\right),
\end{equation}
where $n_k$ is the number of quanta, the first term relates to the quanta emission, and the second one to the absorption.

The mean squared displacement is 
\begin{equation}
\label{A5}
\langle u^2\rangle=\int\frac{d^{3}R}{V}\langle u^2\rangle=\sum_{k}\langle|\vec u_{k}|^2\rangle,
\end{equation}
where $V$ is the volume of the system. Since in our case $n_k=0$, the mean squared displacement has the form 
\begin{equation}
\label{A6}
\langle u^2\rangle=\frac{e^2}{4m^2}\sum_{k}\frac{\langle|\vec{\cal E}_k|^2\rangle}{\omega^{2}_{k}(|\omega_k|+\Omega)^2}\,.
\end{equation}
The energy of zero point oscillations is
\begin{equation}
\label{A7}
\int\frac{d^3R}{4\pi}\langle{\cal E}^2\rangle=\frac{V}{4\pi}\sum_{k}\langle|\vec{\cal E}_k|^2\rangle=\sum_{k}\frac{\hbar|\omega_k|}{2}.
\end{equation}
It follows from here that $\langle|\vec{\cal E}_k|^2\rangle=2\pi\hbar|\omega_k|/V$. Using the summation rule
\begin{equation}
\label{A9}
\sum_{k}=2\int\frac{4\pi k^2dkV}{(2\pi)^3}
\end{equation}
(the coefficient 2 accounts for two photon polarizations) and the relation $\omega_k=ck$, one can obtain from Eq.~(\ref{A6}) \cite{MIGDAL}
\begin{equation}
\label{A10}
\langle u^2\rangle=\frac{r^{2}_{B}}{\pi}\left(\frac{e^2}{\hbar c}\right)^3\int^{\omega_{max}}_{0}\frac{\omega d\omega}{(\omega+\Omega)^2}\,.
\end{equation}
The upper limit $\omega_{max}$ is determined by the condition of non-relativistic motion, that is $\omega_{max}\simeq mc^2/\hbar$. In the 
relativistic region $\vec u_k$ decreases due to the enhancement of the relativistic mass. Using the above rydberg estimate for $\hbar\Omega$ 
one can obtain for the thread radius
\begin{equation}
\label{A11}
r_T=\sqrt{\langle u^2\rangle}=r_c\sqrt{\frac{2e^2}{\pi\hbar c}\ln\frac{\hbar c}{e^2}}\,,
\end{equation}
where $r_c=\hbar/(mc)\simeq 3.86\times 10^{-11}{\rm cm}$ is the electron Compton length. The exact cut off $\Omega$ is not crucial for the 
estimate. The applicability condition of the approach used $\langle u^{\,2}\rangle k^{2}_{max}\sim\langle u^{\,2}\rangle/r^{2}_{c}\ll 1$ is 
valid since $\langle u^{\,2}\rangle/r^{2}_{c}\sim e^2/(\hbar c)$. 
\subsection{Thread properties}
When the wave function is smooth one can consider the electron motion in the effective potential $\langle V(\vec R+\vec u)\rangle$ averaged on
fast electron motions $\vec u$. Expansion up to the second order in $\vec u$ and using Eq.~(\ref{A11}) produces the Lamb shift \cite{MIGDAL} 
coincided (excepting the numerical coefficient) with the exact result \cite{LANDAU3}. 

When the solution of quantum mechanical equations is singular, it is cut off at the distance $r_T$. The thread radius can be estimated from 
Eq.~(\ref{A11}) as $r_T\simeq 0.15\,r_c\simeq 0.58\times 10^{-11}\,{\rm cm}$. 

At distances $r_T<r$ radiation effects are small and one can use the quantum mechanical description of the electron. At $r_c<r$ this the 
Schr\"{o}dinger formalism but at $r_T<r<r_c$ one should apply Dirac equations. In this case, besides the term $\ln r$, the wave function also 
contains the term $r_c/r$ which is small at $r_c\ll r$. See Appendix (\ref{B4}). The total number of electrons is determined by the region
$r\sim r_B$ outside the thread as in a covalent bond. The fraction of electrons in the region $r<r_c$ is small as 
$r^{2}_{c}/r^{2}_{B}\sim (e^2/\hbar c)^3$. 

The wave function is schematically plotted in Fig.~\ref{fig2}(a) where it almost corresponds to the quantum mechanical approach (weak radiation
corrections) outside the thread. Inside the thread the wave function has the peak which can be interpreted as one resulted from a narrow 
potential well. Inside the thread the electron state is coupled to the electromagnetic subsystem and cannot be considered separately. Only the 
total thread energy has a meaning without its separation on electron and electromagnetic ones. 

The exact state, where the electron is coupled to photons, is steady and is characterized by the certain total energy which is conserved. In 
terms of Schr\"{o}dinger equation it would an energy eigenvalue. The state, considered either far from the thread or inside it, corresponds 
to the same energy. The large contribution to the electron kinetic energy $\sqrt{(mc^2)^2+(\hbar c/r_T)^2}-mc^2\simeq 2.93~{\rm MeV}$ inside the 
thread should be compensated by a reduction of the electromagnetic energy. The latter can be interpreted as a potential well at the thread 
region in Fig.~\ref{fig2}(b) instead of the dashed curve indicating the usual Coulomb potential. One can say that the thread state with the 
chemical energy scale $E\sim (1-10)~{\rm eV}$ corresponds to upper energy states in the deep potential well, of the order of 1~MeV, and of the 
radius $r_T$. This well is extended along the thread. 

We emphasize that the interpretation in terms of the potential well is approximate since this is not a single particle quantum mechanics but
coupling to the electromagnetic system.

The reduction of the electromagnetic energy reminds the van der Waals phenomenon when the negative potential between atoms is also formed
by the reduction of the energy of zero point electromagnetic oscillations \cite{CASIMIR,LIF}.

The electron thread energy is continuous and when it is smaller than the lowest level in Fig.~\ref{fig2}(b) the thread state can be destroyed by
quanta absorption only. The high energy parts, involved into thread formation, correspond to the typical time $10^{-22}{\rm s}$. Optical 
processes are slow compared to that time. They lead to an adiabatic motion of the thread parameters but the absorption probability of such quanta
is exponentially small \cite{LANDAU1}. The life time of the thread bond itself is exponentially large (practically infinity) until a high energy 
particle or $\gamma$-quantum destroys it. Also the thread bond cannot be created in chemical or optical processes. The above high energy impact 
is necessary for that. 

The length of the thread bond and a form of the outer electron cloud are determined by chemical mechanisms of interaction with surrounding
electrons. The electron density is mainly localized outside the thread on the usual distance of the order of the Bohr radius as in a covalent bond.
\section{DISCUSSIONS}
\label{disc}
It looks unusual if energies, involved into molecular bonding, are of the order of 1~MeV and the spatial scale is $10^{-3}$ of the size of 
hydrogen atom. At the first sight, it is impossible since chemical bonding is associated with the atomic processes. The point is that the 
electron state can have a tendency to be singular along the line (thread) connected two nuclei in a molecule. This singularity is cut off by 
electron interaction with zero point electromagnetic oscillations. The electron also ``vibrates'' within the thread of the small radius $r_T$ 
where the singularity smears out. 
\begin{figure}
\includegraphics[width=7cm]{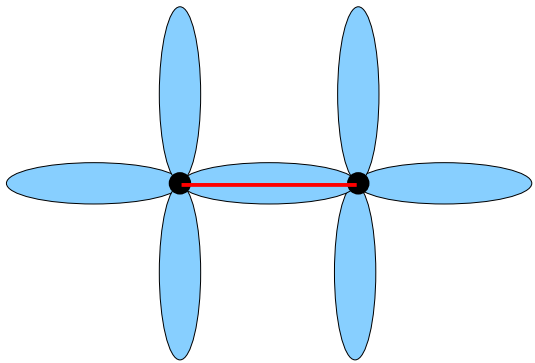}
\caption{\label{fig3}Schematic representation of molecular covalent bonds. The thread bond between two nuclei is shown by the line.}
\end{figure}

The singularity smearing is associated with non-conservation of the electron momentum. The uncertainty of this momentum is 
$\hbar/\sqrt{\langle u^2\rangle}$. In the absence of external potential, as follows from (\ref{A10}), the mean squared displacement is infinite
($\Omega=0$). In this case the uncertainty of the electron momentum is zero and the singularity is not cut off. This also follows from the exact
spectral representation of the electron Green's function in the absence of external fields \cite{LANDAU3}.

The peak of electron wave function in the thread region is associated with formation of the certain deep ($\sim 1~{\rm MeV}$) potential well 
along the thread. The potential well is due to the reduction of the energy of electromagnetic zero point oscillations. The interpretation of 
thread bonding can be in terms of an electron state with the energy $E\sim (1-10)~{\rm eV}$ which corresponds to upper energy states in the 
deep well. The interpretation in terms of the potential well is approximate since this is not a single particle quantum mechanics but coupling 
to the electromagnetic system.

This phenomenon is described by quantum electrodynamics since due to short scales the electron momentum becomes relativistic. This is not the 
unique situation of application of quantum electrodynamics in chemistry. The famous example is van der Waals forces when quantum electrodynamics 
is involved. In that case also the certain steady attraction potential is formed due to the reduction of the energy of electromagnetic zero
point oscillations. 

Since high energy is involved into the thread formation it cannot be created (or destroyed if it exists) by chemical or optical processes. It can 
be created by a high energy particle or $\gamma$-quantum. After that the thread bond lives the exponentially large (practically infinity) time.
A high energy process, of the above type only, can destroy it. 

The schematic representation of the thread bond in a molecule is in Fig.~\ref{fig3}. The thin thread connects two partner sides. The thread bond 
is associated with the outer covalent part where the electron cloud is mainly localized (not on the thread). 

External high energy radiation (natural or artificial) can result in accumulation of thread bonds in biological molecules, for example in DNA. 
One can put a question how thread bonds will affect biological processes, for example, replication of DNA.
\section{CONCLUSIONS}
Unusual chemical bonds are proposed. Each bond is almost covalent but is characterized by the thread of a small radius, 
$\sim 0.6\times 10^{-11}{\rm cm}$, between two nuclei in a molecule. The main electron density is concentrated outside the thread as in a 
covalent bond. The thread is formed by the electron wave function which has a tendency to be singular on it. The singularity along the thread 
is cut off by electron ``vibrations'' due to the interaction with zero point electromagnetic oscillations. The electron energy has its typical 
value of $(1-10)~{\rm eV}$. Due to the small thread radius the uncertainty of the electron momentum inside the thread is large resulting in a 
large electron kinetic energy $\sim 1~{\rm MeV}$. This energy is compensated by formation of a potential well due to the reduction of the energy 
of electromagnetic zero point oscillations. This is similar to formation of a negative van der Waals potential. Thread bonds are stable and 
cannot be created or destructed in chemical or optical processes.

\acknowledgments
I appreciate discussions of related topics with S. J. Brodsky and H. C. Rosu.
\appendix*
\section{SINGULAR SOLUTION OF DIRAC EQUATIONS}
Schr\"{o}dinger equation formally has the solution $\ln r$ in cylindrical coordinates. Below we establish the continuation of this singular 
solution to the region $r<r_c$ where one should use the Dirac formalism. In this case the wave function is bispinor consisting of two spinors
$\varphi$ and $\chi$ \cite{LANDAU3}. Since we are interested by the singular wave functions (large kinetic energy part) one can ignore potential
energy and consider free electron Dirac equations
\begin{equation}
\label{B1}
\left[E_0-i\hbar c(\vec\sigma\nabla)\right]\chi=mc^2\varphi\,,\hspace{0.3cm}\left[E_0+i\hbar c(\vec\sigma\nabla)\right]\varphi=mc^2\chi,
\end{equation}
where $E_0$ is the total relativistic energy and $\vec\sigma$ are the Pauli matrices \cite{LANDAU3}. Substituting the spinor $\chi$ from the 
second equation (\ref{B1}) into the first one we obtain
\begin{equation}
\label{B2}
\left(E^{2}_{0}+\hbar^2c^2\nabla^2\right)\varphi=m^2c^4\varphi.
\end{equation}
The equation for $\chi$ is the same. We use here the relation $(\vec\sigma\nabla)(\vec\sigma\nabla)=\nabla^2$. These equations have solutions 
$N_0(r\sqrt{E^{2}_{0}/c^2-m^2c^2}/\hbar)$ and $(\vec\sigma\nabla)N_0$ where $N_0$ is the Neuman function \cite{GRAD}. 

We use the asymptotic forms $N_0\sim \ln r$ and $\nabla N_0\sim\vec r/r^2$. Those solutions are compatible with the equations (\ref{B1}) if to 
put 
\begin{equation}
\label{B3}
\varphi=\varphi_0\ln r-\frac{ir_c}{2r^2}(\vec r\vec\sigma)\varphi_0,\hspace{0.2cm}\chi=\varphi_0\ln r+\frac{ir_c}{2r^2}(\vec r\vec\sigma)\varphi_0,
\end{equation}
where $\varphi_0$ is some constant spinor. Eqs.~(\ref{B3}) account for the condition that for our non-relativistic energies $(E_0-mc^2)\ll mc^2$, 
two spinors should coincide at large distances $r_c\ll r$ \cite{LANDAU3}. In the standard representation $\Phi=\varphi+\chi$ and 
$\Theta=\varphi-\chi$
\begin{equation}
\label{B4}
\Phi=2\varphi_0\ln r,\hspace{0.2cm}\Theta=-\frac{ir_c}{r^2}(\vec r\vec\sigma)\varphi_0
\end{equation}
at large distances (non-relativistic limit) the wave function is the usual spinor $\Phi$.

\end{document}